\documentclass{revtex4}
\usepackage{amsmath,amsthm}
\usepackage{rotating}
\usepackage{graphicx}

\usepackage{amsfonts}
\newtheorem{theorem}{Theorem}[section]
\newtheorem{lemma}[theorem]{Lemma}
\newtheorem{proposition}[theorem]{Proposition}
\newtheorem{cor}[theorem]{Corollary}

\theoremstyle{remark}
\newtheorem{remark}[theorem]{Remark}

\theoremstyle{definition}
\newtheorem{definition}[theorem]{Definition}

\theoremstyle{example}
\newtheorem{example}[theorem]{Example}

\theoremstyle{notation}

\newcommand{\bra}[1]{\langle#1|}
\newcommand{\ket}[1]{|#1\rangle}

\begin{document}

\title{Grothendieck bound in a single quantum system}            
\author{A. Vourdas}
\affiliation{Department of Computer Science,\\
University of Bradford, \\
Bradford BD7 1DP, United Kingdom\\a.vourdas@bradford.ac.uk}

\begin{abstract}
Grothendieck's bound is used in the context of a single quantum system, in contrast to previous work which used it for multipartite entangled systems and 
the violation of Bell-like inequalities.
Roughly speaking the Grothendieck theorem considers a `classical' quadratic form ${\cal C}$ that uses complex numbers in the unit disc, and takes values less than $1$.
It then proves that if the complex numbers are replaced with vectors in the unit ball of the Hilbert space, then the `quantum' quadratic form ${\cal Q}$   might take values greater than $1$, up to the complex Grothendieck constant $k_G$.
The Grothendieck theorem is reformulated here in terms of arbitrary matrices (which are multiplied with appropriate normalisation prefactors), so that it is directly applicable to quantum quantities.
The emphasis in the paper is in the `Grothendieck region' $(1,k_G)$, which is a classically forbidden region in the sense that  
${\cal C}$ cannot take values in it.
Necessary (but not sufficient) conditions  for ${\cal Q}$ taking values in the Grothendieck region are given.
 Two examples that involve physical quantities in systems with $6$ and $12$-dimensional Hilbert space, are shown to lead to ${\cal Q}$ in the Grothendieck region $(1,k_G)$.
They involve projectors of the overlaps of novel generalised coherent states that resolve the identity and have a discrete isotropy.
\end{abstract}
\maketitle

\section{Introduction}
Various inequalities play an important role in a quantum context.
Uncertainty relations, entropic inequalities \cite{M1,M2,M3}, Bell inequalities in multipartite entangled systems (e.g., \cite{C1,C2,C3}), etc. 
An important inequality in a pure mathematics context is the Grothendieck theorem \cite{GR1,GR2,GR3,GR4},
which has been applied in many areas.
Work on examples that take values near the real and complex Grothendieck constants, is discussed in \cite{GR5,GR6,GR7}.
Applications to graph theory and computer science are discussed in \cite{GG1}.

The  Grothendieck theorem has been used in a quantum context, for multipartite entangled systems \cite{E1,E2,E3,E4,E5,E6,E7,E8}.
Roughly speaking violation of Bell-like inequalities (e.g., the Clauser, Horne, Shimony, and Holt inequality \cite{C2}), 
corresponds to the fact that in the Hilbert space formalism some quantities can take values greater than one and up to the 
Grothendieck constant $k_G$.
We note that in a quantum context, all previous work on the Grothendieck theorem is for entangled multipartite systems.
In this paper we discuss a different application of the Grothendieck theorem which involves a single quantum system (and is not related to entanglement).

The original formulation of the Grothendieck theorem \cite{GR1} was in the context of tensor product of Banach spaces,
and this leads to the impression that applications in a quantum context should be for multipartite systems described by tensor products of Hilbert spaces.
But all later mathematical work on Grothendieck's theorem \cite{GR2,GR3,GR4} emphasised that the theory can also be formulated and is interesting, outside the framework of tensor product theory.
And this motivated our work that uses a single quantum system (and is totally unrelated to tensor products, multipartite systems and entanglement).

The Grothendieck theorem considers a quadratic form ${\cal C}$ that uses complex numbers in the unit disc, and takes values in $[0,1]$.
It then proves that if the complex numbers are replaced with vectors in the unit ball of the Hilbert space, the corresponding quantity  ${\cal Q}$ 
takes values in $[0,k_G)$, where $k_G$ is the complex Grothendieck constant (for which it is known that $1<k_G\le 1.4049$).
The complex numbers in the unit disc can be interpreted as classical quantities, and the vectors as quantum quantities 
(which embody the superposition principle through addition of vectors). 
In this sense going from  ${\cal C}$ to ${\cal Q}$ is a passage from classical to quantum mechanics.

The Grothendieck bound is a `mathematical ceiling' of the Hilbert formalism, and in turn of the quantum formalism which is intimately related to it.
The `classical ceiling' for ${\cal C}$ is $1$, while the `quantum ceiling' for  ${\cal Q}$ is $k_G$.
When ${\cal Q}$ crosses from $[0,1]$ to the `Grothendieck region' $(1,k_G)$, it enters a classically forbidden region, and the study of this is the objective of this paper because it is `quantum mechanics on the edge'.
There are large families of examples which show quantum behaviour (e.g., the Wigner and Weyl functions show clearly quantum interference), and yet  the corresponding ${\cal Q} \in [0,1]$ (section \ref{PPP}).
In this sense the Grothendieck region is new territory of quantum mechanics with rare examples, e.g., the examples in sections 7, 8 (in Mathematics sometimes the interesting features are found in rare examples).
Future work with random sampling of matrices, could quantify how rare the examples with ${\cal Q}\in(1,k_G)$ are.

We reformulate the Grothendieck theorem in terms of arbitrary $d\times d$ complex matrices, which are normalised.
In this form the Grothendieck theorem is directly applicable to quantum mechanical quantities which are usually products of matrices.
In order to do this we need two different types of normalisation which are introduced in sections 3,4.
We then reformulate the Grothendieck theorem in terms of arbitrary matrices, which are normalised with these factors (proposition \ref{GRO}), in section 5.

In section 6 we show that there are large families of examples with ${\cal Q}$ which cannot take values in the Grothendieck region $(1,k_G)$.
This leads to necessary (but not sufficient) conditions  for ${\cal Q}$ taking values in the Grothendieck region
(corollary \ref{GRO10} and proposition \ref{pro38}).

In section 7 we give a physical example that leads to  ${\cal Q}\in (1,k_G)$. 
This example involves a projector in a quantum system with $6$-dimensional Hilbert space.
This projector contains the overlaps of  novel generalised coherent states in a $3$-dimensional Hilbert space, that resolve the identity and have a discrete isotropy.
An analytical method shows that ${\cal Q}$ enters the Grothendieck region $(1,k_G)$, and a numerical maximisation method 
shows that in this example, ${\cal Q}$ can take the maximum value $\frac{6}{5}$.

In section 8 we generalise this example.
Starting from novel generalised coherent states in a $4$-dimensional Hilbert space that resolve the identity and have a discrete isotropy,
we define (through their overlaps) a projector in a $12$-dimensional Hilbert space, which leads to ${\cal Q}\in (1,k_G)$.

We conclude in section 9 with a discussion of our results.

\section{Preliminaries and notation}

We consider a quantum system with variables in ${\mathbb Z}_d$ (the integers modulo $d$), described by a $d$-dimensional Hilbert space $H(d)$.
For a vector ${\bf x}$ in $H(d)$, the scalar product and the norm  are given by
\begin{eqnarray}\label{478}
({\bf x},{\bf y})=\sum_ix_i^*y_i;\;\;\;||{\bf x}||=\sqrt{({\bf x},{\bf x})}.
\end{eqnarray}
We will also use the usual bra and ket notation for normalised vectors in $H(d)$.

Let $M$ be a complex $d\times d$ matrix with singular values $s_i$. We define the matrix norms
\begin{eqnarray}
||M||_1=\sum _{i,j} |M_{ij}|;\;\;\;||M||_2=\sqrt{\sum_{i,j}|M_{ij}|^2}=\sqrt{{\rm Tr}(MM^\dagger)}=\sqrt{\sum _is_i^2}
\end{eqnarray}
$||M||_2$ is the Frobenius norm, and it is invariant under unitary transformations.
If $e_i$ are the eigenvalues of $M$, its spectral radius is
\begin{eqnarray}
e_{\rm max}=\max \{|e_i|\};\;\;\;e_{\rm max}\le ||M||_1;\;\;\;e_{\rm max}\le ||M||_2.
\end{eqnarray}
We also define the spectral norm:
\begin{eqnarray}
||M||_{\rm spe}=\max _{\bf x}\frac{||M{\bf x}||}{||{\bf x}||}={\mathfrak s}_{\rm max}.
\end{eqnarray}
Here ${\mathfrak s}_{\rm max}$ is the largest singular value of $M$ (for normal matrices ${\mathfrak s}_{\rm max}=e_{\rm max}$).

\begin{example}
We consider the Fourier matrix
\begin{eqnarray}\label{4}
F_{ij}=\frac{1}{\sqrt{d}}\omega ^{ij};\;\;\;\omega=\exp \left( \frac{i2\pi}{d} \right);\;\;\;F^4={\bf 1}.
\end{eqnarray}
In this case
\begin{eqnarray}
||F||_1=d\sqrt{d};\;\;\;||F||_2=\sqrt{d};\;\;\;e_{\rm max}=1.
\end{eqnarray}
\end{example}
\begin{example}\label{ex39}
We consider a density matrix $\rho$. In this case
\begin{eqnarray}\label{RRR}
||\rho||_1\ge 1;\;\;\;1\ge e_{\rm max}\ge \frac{1}{d}.
\end{eqnarray}
Indeed 
 $\rho=UDU^\dagger$, where $U$ is a unitary matrix and $D$ is a diagonal matrix with its eigenvalues $e_i$.
Then
\begin{eqnarray}
||\rho||_1\ge \sum _{i}|\rho_{ii}|= \sum_{i}|\sum_k U_{ik}e_kU^*_{ik}|\ge |\sum_k e_k(\sum _iU_{ik}U^*_{ik})|=1.
\end{eqnarray}
Also for non-negative numbers the maximum value is greater than the average value which for the eigenvalues of a density matrix is $\frac{1}{d}$.
\end{example}

\subsection{The symmetric group $\Sigma_\nu$}
Let    $\pi$ be a permutation of  elements of ${\mathbb Z}_\nu$:
\begin{eqnarray}\label{per}
(0,1,...,\nu-1)\;\overset{\pi} \rightarrow\;(\pi(0),\pi(1),...,\pi(\nu-1)).
\end{eqnarray}
$\pi$ with the composition 
\begin{eqnarray}
(0,1,...,\nu-1)\;\overset{\pi} \rightarrow\;(\pi(0),\pi(1),...,\pi(\nu-1))\overset{\varpi} \rightarrow\;(\varpi [\pi(0)],\varpi[\pi(1)],...,\varpi[\pi(\nu-1)]).
\end{eqnarray}
 is an element of the symmetric group $\Sigma_\nu$\cite{sym}  which has $\nu!$ elements.

A permutation matrix $\tau_\pi$ is a $\nu\times \nu$ matrix with elements 
\begin{eqnarray}\label{PPQ}
\tau_\pi(i,j)=\delta(\pi(i),j),
\end{eqnarray}
Each row and each column have one element equal to $1$ and the other $\nu-1$ elements equal to $0$.

\section{The set ${\cal S}_d$ of $d\times d$ matrices}

In this section we introduce the first normalisation of $d\times d$ complex matrices.
We regard a matrix $M$ as a set of $d$ row vectors which we denote as $\widehat {M_i}$, with norms 
\begin{eqnarray}\label{5}
||\widehat {M_i}||=\sqrt{\sum _j |M_{ij}|^2}=\sqrt{(MM^\dagger)_{ii}};\;\;\;
\sum_i||\widehat {M_i}||^2=(||M||_2)^2.
\end{eqnarray}
We define the normalisation factor for the matrix $M$
\begin{eqnarray}\label{109}
{\cal N}(M)=\max_i ||\widehat {M_i}||=\max_i \sqrt{\sum _j |M_{ij}|^2}=\max_i \sqrt{(MM^\dagger)_{ii}}.
\end{eqnarray}
If $z$ is a complex number, then
\begin{eqnarray}
{\cal N}(zM)=|z|{\cal N}(M).
\end{eqnarray}

In general 
\begin{eqnarray}\label{9}
{\cal N}(UMU^\dagger)\ne {\cal N}(M)
\end{eqnarray}
where $U$ is a unitary transformation. 
Therefore in the calculation of the normalisation factor of an operator $\theta$,
it is important to specify an orthonormal basis $\ket{u_i}$ with respect to which 
we first find the matrix $\bra{u_i}\theta\ket{u_j}$ and then the corresponding ${\cal N}(\theta;\ket{u_i})$.
In a different basis the ${\cal N}$ will be different (in general).
For operators, the notation will indicate clearly the basis.

In general ${\cal N}(M)\ne{\cal N}(M^\dagger)$.
But for normal matrices ${\cal N}(M)={\cal N}(M^\dagger)$, as shown in the following proposition.

\begin{proposition}\label{pro12}
\mbox{}
\begin{itemize}
\item[(1)]
If $M$ is a matrix with singular values ${\mathfrak s}_i$ 
\begin{eqnarray}\label{8}
 \frac{1}{\sqrt d}||M||_2=\sqrt{\frac{\sum_i{\mathfrak s}_i^2}{d}}\le {\cal N}(M)\le ||M||_2
\end{eqnarray}
Under unitary transformations the two bounds are invariant, and the 
${\cal N}(M)$ varies between them.
 If all $||\widehat {M_i}||$ are equal to each other, then the left inequality becomes equality.
If all $||\widehat {M_i}||$ except one are zero (i.e., if all rows except one are zero), then the right inequality becomes equality.

\item[(2)]
If $M$ is a normal matrix and $e_{\rm max}$ its spectral radius, then 
\begin{eqnarray}\label{BB1}
\frac{1}{\sqrt{d}}||M||_2=\sqrt{\frac{\sum_i|e_i|^2}{d}}\le {\cal N}(M)={\cal N}(M^\dagger)\le e_{\rm max}.
\end{eqnarray}
Under unitary transformations the two bounds are invariant, and the 
${\cal N}(M)$ varies between them.
 If all $||\widehat {M_i}||$ are equal to each other, then the left inequality becomes equality.
If $M$ is diagonal, then the right inequality becomes equality.

In the special case that $M$ is a density matrix $\rho$, Eq.(\ref{BB1}) becomes 
\begin{eqnarray}\label{BB0}
\sqrt{\frac{{\rm Tr}(\rho^2)}{d}}\le {\cal N}(\rho)\le e_{\rm max}\le 1.
\end{eqnarray}
\item[(3)]
For a unitary matrix $U$, we get ${\cal N}(U)=1$ and $||U||_2=\sqrt{d}$.

\end{itemize}
\end{proposition}
\begin{proof}
\mbox{}
\begin{itemize}
\item[(1)]
Since $\sum_i||\widehat {M_i}||^2=(||M||_2)^2$, it follows that the maximum value of $||\widehat {M_i}||$ is $||M||_2$.
If all $||\widehat {M_i}||$ except one are zero, then clearly the maximum value of $||\widehat {M_i}||$ is equal to $||M||_2$.

The $\sum_i||\widehat {M_i}||^2=(||M||_2)^2$ also implies that the average value of $||\widehat {M_i}||$ is $\frac{1}{\sqrt d}||M||_2$.
For positive numbers, the maximum value is greater than the average value and this proves the left inequality in Eq.(\ref{8}).
 If all $||\widehat {M_i}||$ are equal to each other, then the maximum value of $||\widehat {M_i}||$ is equal to their average. In this case the left inequality becomes equality.

\item[(2)]
We have
\begin{eqnarray}\label{BB2}
{\cal N}(M)=\max_i \sqrt{(MM^\dagger)_{ii}};\;\;\;{\cal N}(M^\dagger)=\max_i \sqrt{(M^\dagger M)_{ii}}
\end{eqnarray}
For normal matrices $M^\dagger M=MM^\dagger $ and therefore ${\cal N}(M)={\cal N}(M^\dagger)$.

A normal matrix can be diagonalised with a unitary transformation as $M=UDU^\dagger$, where $U$ is a unitary matrix, and $D={\rm diag}(e_0,..., e_{d-1})$.
Then
\begin{eqnarray}\label{PP}
\sum _j|M_{ij}|^2&=&\sum_j|\sum_k U_{ik}e_kU^*_{jk}|^2=\sum_j\left (\sum_{k_1,k_2}U_{ik_1}e_{k_1}U^*_{jk_1}U_{ik_2}^*e_{k_2}^*U_{jk_2}\right )
\nonumber\\&=&\sum_{k_1,k_2}U_{ik_1}e_{k_1}U_{ik_2}^*e_{k_2}^* \left (\sum _j  U^*_{jk_1}U_{jk_2}\right )=\sum_{k_1,k_2}U_{ik_1}e_{k_1}U_{ik_2}^*e_{k_2}^*\delta_{k_1k_2}
\nonumber\\&=&\sum_{k_1}|U_{ik_1}e_{k_1}|^2\le (e_{\rm max})^2\sum_{k_1}|U_{ik_1}|^2=(e_{\rm max})^2
\end{eqnarray}
This proves the right inequality in Eq.(\ref{BB1}). 
Clearly if $M$ is diagonal, this inequality becomes equality.

The left inequality in Eq.(\ref{BB1}) is the same as in Eq.(\ref{8}).
We note here that for normal matrices $||M||_2=\sqrt{\sum _i|e_i|^2}$.

For a density matrix, $||\rho||_2=\sqrt{{\rm Tr}(\rho^2)}=\sqrt{\sum _ie_i^2}$.

\item[(3)]
From Eq.(\ref{109}) follows that ${\cal N}(U)=1$ and then $||U||_2=\sqrt{d}$.

\end{itemize}
\end{proof}

\begin{definition}
${\cal S}_d$ is the set of all matrices $M$ with ${\cal N}(M)\le1$.
\end{definition}
If $V$ belongs to ${\cal S}_d$, then the matrix $zV$ with $|z|\le 1$ also belongs to ${\cal S}_d$.

\begin{proposition}\label{pro00}
\mbox{}
\begin{itemize}
\item[(1)]
For any $d\times d$ complex matrix $\cal V$, the matrix
\begin{eqnarray}\label{777}
V=\frac{1}{{\cal N}({\cal V})}{\cal V},
\end{eqnarray}
belongs to ${\cal S}_d$.

\item[(2)]
All unitary matrices belong to ${\cal S}_d$. In particular the matrix ${\bf 1}$ belongs to ${\cal S}_d$.

\item[(3)]
If $V$ belongs to ${\cal S}_d$, the $V^\dagger$ might not belong to ${\cal S}_d$.
\end{itemize}
\end{proposition}

\begin{proof}
\mbox{}
\begin{itemize}
\item[(1)]
The normalisation factor ensures that the $d$ rows of $V$, are vectors with norm less or equal to $1$. In fact the maximum of these $d$ norms is one.

\item[(2)]
The normalisation factor for unitary matrices is one.
\item[(3)]

We have explained earlier that in general ${\cal N}(V)\ne {\cal N}(V^\dagger)$, and therefore the ${\cal N}(V)\le 1$ does not imply ${\cal N}(V^\dagger)\le 1$.
\end{itemize}
\end{proof}

\section{The set $G_d$ of $d\times d$ matrices}\label{sec35}
 
 In this section we introduce the second normalisation of $d\times d$ complex matrices.
 Let ${\cal D}=\{|z|\le 1\}$ be the unit disc in the complex plane, and ${\cal D}^d$ the set of all $d$-tuples $(s_0,...,s_{d-1})$ where $s_i\in\cal D$.
 
 \begin{definition}
$G_d$ is the set of all  $d\times d$ complex matrices $\theta$, such that for all $d$-tuples
$(s_0,...,s_{d-1})$ and $(t_0,...,t_{d-1})$ in ${\cal D}^d$, we get
\begin{eqnarray}\label{892}
{\cal C}=\left |\sum_{i,j}\theta _{ij}s_it_j\right |\le 1;\;\;\;|s_i|\le 1;\;\;\;|t_j|\le 1.
\end{eqnarray}
\end{definition}
If $\theta$ belongs in $G_d$, then the matrix $z\theta$ with $|z|\le1$ also belongs to $G_d$. 

\begin{definition}
$G_d^\prime$ is the set of all $d\times d$ complex matrices $\theta$, such that for all $d$-tuples
$(s_0,...,s_{d-1})$ and $(t_0,...,t_{d-1})$ where $\sum_i|s_i|^2\le d$, $\sum _j|t_j|^2\le d$, we get
\begin{eqnarray}\label{892A}
\left |\sum_{i,j}\theta _{ij}s_it_j\right |\le 1;\;\;\;\sum_i|s_i|^2\le d;\;\;\;\sum _j|t_j|^2\le d.
\end{eqnarray}
\end{definition}
Here we use $d$-tuples which belong to a superset of  ${\cal D}^d$, and therefore $G_d^\prime\subset G_d$.
The physical importance of $G_d^\prime$ is seen in proposition \ref{pro38} below.
 For normal matrices $\theta\in G_d^\prime$ the quantity ${\cal Q}$ (defined later) cannot take values in the Grothendieck region $(1,k_G)$.
A necessary (but not sufficient) condition for ${\cal Q}\in (1,k_G)$ is that $\theta\in G_d\setminus G_d^\prime$.

\begin{definition}
For any $d\times d$ complex matrix $\theta$, 
\begin{eqnarray}\label{29}
g(\theta)=\sup \left \{\left |\sum_{i,j}\theta_{ij} s_it_j\right |;\;\;\;|s_i|\le 1;\;\;\;|t_j|\le 1\right \},
\end{eqnarray}
and
\begin{eqnarray}\label{23}
g^\prime (\theta)=\sup \left \{\left |\sum_{i,j}\theta_{ij} s_it_j\right |;\;\;\;\sum _i|s_i|^2\le d;\;\;\;\sum _j|t_j|^2\le d\right \}.
\end{eqnarray}
\end{definition}
\begin{proposition}
$g^\prime (\theta)=d{\mathfrak s}_{\rm max}$ where ${\mathfrak s}_{\rm max}$ is the largest singular value of $\theta$.
For normal matrices $g^\prime (\theta)=de_{\rm max}$, and for unitary matrices $g^\prime (\theta)=d$.
\end{proposition}
\begin{proof}
We consider the vectors 
\begin{eqnarray}
{\bf s}=\left (\frac{s_i}{\sqrt{d}}\right);\;\;\;\theta {\bf t}=\left (\sum _j\theta_{ij}\frac{t_j}{\sqrt{d}}\right ).
\end{eqnarray}
In Eq.(\ref{23}) we have $\sum _i|s_i|^2\le d$ and $\sum _j|t_j|^2\le d$, but
since we are interested in the supremum we take $||{\bf s}||=||{\bf t}||= 1$.
Then
\begin{eqnarray}
||\theta||_{\rm spe}=\max_{||\bf t||=1}||\theta {\bf t}||={\mathfrak s}_{\rm max}.
\end{eqnarray}
We use the Cauchy-Schwartz inequality and we get
\begin{eqnarray}\label{A25}
\left |\sum_{i,j}\theta_{ij} s_it_j\right |={d}|({\bf s},\theta {\bf t})|\le {d}||{\bf s}||\cdot||\theta{\bf t}||\le d{\mathfrak s}_{\rm max}.
\end{eqnarray}
We note that we get equality only if the two vectors are proportional to each other, i.e., if we choose
\begin{eqnarray}\label{A26}
{\bf s}=\frac{\theta{\bf t}}{||\theta{\bf t}||}.
\end{eqnarray}
This proves that $g^\prime (\theta)=d{\mathfrak s}_{\rm max}$.

For normal matrices ${\mathfrak s}_{\rm max}=e_{\rm max}$, and for unitary matrices $e_{\rm max}=1$.

\end{proof}

It is easily seen that
\begin{itemize}
\item
By definition if $\theta \in G_d$ then $g(\theta)\le 1$, and if $\theta \in G_d^\prime$ then $g^\prime (\theta)\le 1$.
\item
The following inequalities hold:
\begin{eqnarray}\label{27}
g(\theta)\le g^\prime (\theta)=d{\mathfrak s}_{\rm max};\;\;\;g(\theta)\le||\theta||_1.
\end{eqnarray}

\item
For any $d\times d$ complex matrix $\theta$, we get
\begin{eqnarray}\label{25}
&&\lambda\le \frac{1}{d{\mathfrak s}_{\rm max}} \;\rightarrow\;\lambda\theta \in G_d^\prime\nonumber\\
&&\frac{1}{d{\mathfrak s}_{\rm max}} \le \lambda\le \frac{1}{g(\theta)}\;\rightarrow\;\lambda \theta \in G_d\setminus G_d^\prime.
\end{eqnarray}

\item
Necessary (but not sufficient) conditions for $\theta\in G_d^\prime$, are:
\begin{eqnarray}\label{RR}
|\theta_{ij}|\le \frac{1}{d};\;\;\;||\theta||_1\le d;\;\;\;||\theta||_2\le 1.
\end{eqnarray}
We prove this  if we take $s_i=t_j=\sqrt{d}$ and the rest of them zero.

\item
Necessary (but not sufficient) conditions for $\theta\in G_d$, are
\begin{eqnarray}
|\theta_{ij}|\le 1;\;\;\;||\theta||_1\le d^2;\;\;\;||\theta||_2\le d.
\end{eqnarray}
We prove this if we take $s_i=t_j=1$ and the rest of them zero.

\item
If $U$ is a unitary matrix then in general
\begin{eqnarray}\label{341}
g (U\theta U^\dagger)\ne g(\theta)
\end{eqnarray}

\end{itemize}

\begin{example}
We consider the  matrix 
\begin{eqnarray}
&&\theta_{ij}=1\;{\rm if}\;i=a\;{\rm and}\;j=b\nonumber\\
&&\theta_{ij}=0\;{\rm otherwise}
\end{eqnarray}
Then:
\begin{itemize}
\item
For $\frac {1}{d} < \lambda \le 1$, the matrix $\lambda \theta$ belongs in $G_d$ but does not belong in $G_d^\prime$.
\item
For $\lambda \le \frac {1}{d}$ the matrix $\lambda \theta$ belongs in $G_d^\prime$ (and therefore in $G_d$).
\end{itemize}
Indeed, if $|s_i|\le 1$, $|t_j|\le 1$ and $\lambda \le 1$ the matrix $\lambda \theta$ gives
\begin{eqnarray}
\lambda \left |\sum_{i,j}\theta _{ij}s_it_j\right |=\lambda |s_at_b|\le 1.
\end{eqnarray}
But if $s_i=t_j=\sqrt{d}$ and the other $s_k, t_k$ are zero (in which case $\sum_i|s_i|^2\le d$ and $\sum _j|t_j|^2\le d$), we get
\begin{eqnarray}
\lambda \left |\sum_{i,j}\theta _{ij}s_it_j\right |=\lambda d.
\end{eqnarray}
Therefore only for $\lambda\le \frac{1}{d}$ the matrix $\lambda \theta \in G_d^\prime$.
And for $\frac {1}{d} < \lambda \le 1$, the matrix $\lambda \theta$ belongs in $G_d\setminus G_d^\prime$.
\end{example}

\begin{example}\label{ex34}
Let  $\pi$ be a permutation as in Eq.(\ref{per}).
We consider the $d\times d$ matrix with elements 
\begin{eqnarray}\label{PP}
\theta_{ij}=a_i\delta(i,\pi(j));\;\;\;||\theta||_1=\sum _i|a_i|\le 1,
\end{eqnarray}
They are related to the permutation matrices in Eq(\ref{PPQ}), but here the ones are replaced with $a_i$ (with $\sum _i|a_i|\le 1$).
Each row and each column have one of the elements $a_i$, and the other $d-1$ elements equal to $0$.
Then $g(\theta)=||\theta||_1\le 1$ and $\theta$ belongs to $G_d$.

Examples are the `backslash' matrices
\begin{eqnarray}
\theta_{ij}=a_i\delta_{i,j+k};\;\;\;;\;\;\;||\theta||_1=\sum _i|a_i|\le 1,
\end{eqnarray}
and also the `forward slash' matrices 
\begin{eqnarray}
\theta_{ij}=a_i\delta_{i,-j+k};\;\;\;||\theta||_1=\sum _i|a_i|\le 1.
\end{eqnarray}
\end{example}

We have seen that $g(\theta)\le ||\theta||_1$, and we now explore when this is equality and when it is a strict inequality.
We consider the expression $\sum_{i,j}\theta_{ij} s_it_j$
and express the non-zero elements of $\theta_{ij}$ and the $s_i, t_j$ as
\begin{eqnarray}
\theta_{ij}=|\theta_{ij}|\exp(i\phi _{ij});\;\;\;s_i=|s_i|\exp(-i\chi_i);\;\;\;t_j=|t_j|\exp(-i\psi_j).
\end{eqnarray}
Then
\begin{eqnarray}
\left |\sum _{i,j}\theta_{ij}s_it_j\right |&=&\left |\sum _{i,j}|\theta_{ij}s_it_j|\exp[i(\phi _{ij}-\chi_i-\psi_j)]\right |
\end{eqnarray}
Therefore if for all $\theta_{ij}\ne 0$ we have
\begin{eqnarray}\label{GG}
\phi _{ij}-\chi_{i}-\psi_{j}=0,
\end{eqnarray}
then $g(\theta)$ takes its maximum possible value $||\theta||_1$ (we take $|s_i|=|t_j|=1$, and choose the solutions of the system as their phases $\chi_i, \psi_j$). 
If $N$ elements $\theta_{ij}$ are non-zero, this is a system of $N$ equations with $2d$ unknowns ($\chi_i, \psi_j$).
If this system has a solution then $g(\theta)= ||\theta||_1$, otherwise $g(\theta)< ||\theta||_1$.

Roughly speaking the system will have a solution when $N\le 2d$, and it will have no solution when $N>2d$.
But this is not always true (see example \ref{ex22} below).

\begin{lemma}\label{lem1}
We express the system in Eq.(\ref{GG}) as $AB=C$ where $A$ is a $N\times (2d)$ matrix, and $B,C$ are columns with $2d$ and $N$ elements, correspondingly.
We also consider the $N\times (2d+1)$ `augmented matrix' $D$, which is the matrix $A$ with the matrix $C$ as an extra column.
Then
\begin{itemize}
\item
if ${\rm rank}(D)>{\rm rank}(A)$, then the system has no solution and $g(\theta)<||\theta||_1$.
\item
if ${\rm rank}(D)={\rm rank}(A)$, then the system has a solution and $g(\theta)=||\theta||_1$.
\end{itemize}
\end{lemma}
\begin{proof}
This is based on the Rouche-Capelli theorem.
\end{proof}

In the proposition below we give general examples of matrices in $G_d$. We also give stronger results (i.e., with larger prefactor) for normal matrices.

\begin{proposition}\label{pro98}
\begin{itemize}
\item[(1)]
For any $d\times d$ complex matrix $M$, the matrix $\lambda M$ with
\begin{eqnarray}\label{777}
\lambda\le \frac{1}{||M||_1}
\end{eqnarray}
belongs to $G_d$. 
\item[(2)]
Let $M$ be a $d\times d$ normal matrix with spectral radius $e_{\rm max}$.
 Then 
\begin{eqnarray}\label{43}
&&{\lambda}M\in G_d^\prime;\;\;\;{\rm for}\;\;  \lambda\le \frac{1}{d e_{\rm max}}\nonumber\\
&&\lambda M \in G_d\setminus G_d^\prime;\;\;\;{\rm for}\;\; \frac{1}{de_{\rm max}} \le \lambda\le \frac{1}{g(M)}.
\end{eqnarray}

\end{itemize}
\end{proposition}
\begin{proof}
\begin{itemize}
\item[(1)]
This follows from the fact that $g(M)\le ||M||_1$.
\item[(2)]
We use Eq.(\ref{25}) with ${\mathfrak s}_{\rm max}=e_{\rm max}$ for normal matrices.

\end{itemize}
\end{proof}

\begin{example}\label{ex22}
We consider the Hermitian matrix
\begin{eqnarray}\label{46}
M=\begin{pmatrix}
a&b\\
b^*&-c
\end{pmatrix};\;\;\;a\ge c> 0;\;\;\;b\in{\mathbb C};\;\;\;b\ne 0.
\end{eqnarray}
In this case the system of Eq.(\ref{GG}) has $4$ equations with $4$ unknowns:
\begin{eqnarray}
&&\chi_{0}+\psi_{0}=0\nonumber\\
&&\chi_{1}+\psi_{1}=-\pi\nonumber\\
&&\chi_{0}+\psi_{1}=\arg(b)\nonumber\\
&&\chi_{1}+\psi_{0}=-\arg(b),
\end{eqnarray}
It is easily seen that  this system has no solution, and therefore $g(M)<||M||_1=a+c+2|b|$ (for more complicated examples we can use lemma \ref{lem1}).

For the matrix $M$ we find
\begin{eqnarray}
e_{\rm max}=\frac{1}{2}\left [a-c+\sqrt{(a+c)^2+4|b|^2}\right ].
\end{eqnarray}
Then
\begin{eqnarray}\label{PPB1}
&&{\lambda}M\in G_2^\prime\;\;\;{\rm for}\;\;\lambda\le \frac{1}{ \left [a-c+\sqrt{(a+c)^2+4|b|^2}\right ]}\nonumber\\
&&{\lambda}M\in G_2\setminus G_2^\prime\;\;\;{\rm for}\;\;\frac{1}{ \left [a-c+\sqrt{(a+c)^2+4|b|^2}\right ]}\le \lambda\le \frac{1}{a+c+2|b|}
\end{eqnarray}
Here we do not know the exact value of $\frac{1}{g(M)}$, and we used the $\frac{1}{a+c+2|b|}$ which is a lower bound to it.
But clearly there are some values of $\lambda$ greater than $\frac{1}{a+c+2|b|}$ for which ${\lambda}M\in G_2\setminus G_2^\prime$.

\end{example}

\section{The Grothendieck theorem in terms of matrices in ${\cal S}_d$ and $G_d$}

 Let ${\cal B}_d$ the unit ball in the Hilbert space $H(d)$ (it contains vectors $\lambda \ket{u}$ with $\lambda\le 1$).
The  Grothendieck theorem proves that if Eq.(\ref{892}) holds (i.e., if $\theta \in G_d$), then for all $(\lambda_0\ket{u_0},...,\lambda_{d-1}\ket{u_{d-1}})$ and $(\mu_0\ket{v_0},...,\mu_{d-1}\ket{v_{d-1}})$ where $\lambda _i\ket{u_i}, \mu_j\ket{v_j}\in{\cal B}_d$, we get
\begin{eqnarray}\label{344}
{\cal Q}=\left |\sum_{i,j}\theta_{ij}\lambda_i\mu_j\bra{u_i}v_j\rangle \right |\le k(d)\le k_G;\;\;\;\lambda_i, \mu_j\le1.
\end{eqnarray}
Here $k(d)$ is a constant that depends on the dimension $d$.
$k(d)$ is a non-decreasing function of $d$ (because every matrix can be written as a larger matrix with extra zeros). Then
\begin{eqnarray}
\lim _{d\to \infty}k(d)=k_G.
\end{eqnarray}
$k_G$ is the complex Grothendieck constant, which does not depend on the dimension $d$. Its exact value is not known, but it is known that $1<k_G\le 1.4049$.

An obvious upper bound for the left hand side in Eq.(\ref{344}) is $||\theta||_1$. Therefore we rewrite Eq.(\ref{344}) as
\begin{eqnarray}\label{34}
{\cal Q}=\left |\sum_{i,j}\theta_{ij}\lambda_i\mu_j\bra{u_i}v_j\rangle \right |\le \min (k_G, ||\theta||_1);\;\;\;\lambda_i, \mu_j\le1.
\end{eqnarray}

\begin{proposition}\label{GRO}
Let $\theta, V, W$ be $d\times d$ matrices such that $\theta \in G_d$ and $W,V\in {\cal S}_d$. Then 
\begin{eqnarray}\label{900}
{\cal Q}=|{\rm Tr}(\theta VW^\dagger )|\le  \min( k_G, ||\theta||_1).
\end{eqnarray}
For arbitrary matrices, appropriate normalisation (as in propositions \ref{pro00}, \ref{pro98}) leads to matrices that satisfy the requirement $\theta \in G_d$ and $W,V\in {\cal S}_d$.
\end{proposition}
\begin{proof}
We consider  the $d\times d$ matrix $A$ with elements
\begin{eqnarray}
A_{ji}=\lambda_i\mu_j\bra{u_i}v_j\rangle;\;\;\;\lambda_i,\mu_j\le 1.
\end{eqnarray}
The matrix $A$ can be written as $A=VW^\dagger $ where $V$ is a $d\times d$ matrix that has the components of $\mu_j\ket{v_j}$ in the $j$-row, and $W$ is a matrix that has the components of $\lambda_i\ket{u_i}$ in the $i$-row
(therefore $W^\dagger$ has the complex conjugates of the components of $\lambda_i\ket{u_i}$ in the $i$-column).
Consequently $W,V$ are matrices with $d$ row vectors that have norm less or equal to $1$:
\begin{eqnarray}\label{QQ}
\sum_j|W_{ij}|^2\le 1;\;\;\;\sum_j|V_{ij}|^2\le 1.
\end{eqnarray}
Therefore $W,V$ belong to ${\cal S}_d$. So Eq.(\ref{34}) leads to Eq.(\ref{900}).

If we take $\lambda_i=1$ and $\ket{u_i}$ to be the orthonormal basis 
\begin{eqnarray}\label{ortho}
\ket{u_i}=(0,...,0,1_i,0...,0)^\dagger
\end{eqnarray}
then $W={\bf 1}$.
\end{proof}

From this proposition follows immediately the following:
\begin{cor}\label{GRO10}
Let $\theta, V, W$ be $d\times d$ matrices such that $\theta \in G_d$  (and therefore by definition $g(\theta)\le 1$) and $W,V\in {\cal S}_d$. 
A necessary (but not sufficient) condition for ${\cal Q}=|{\rm Tr}(\theta VW^\dagger )|\in(1,k_G)$ is that
\begin{eqnarray}\label{cond}
g(\theta)\le 1<||\theta||_1.
\end{eqnarray}

\end{cor}

\subsection{Non-diagonal elements are important for   ${\cal Q}\in(1,k_G)$}\label{sec12}

\begin{proposition}
Let $W,V\in {\cal S}_d$ and $\theta$ be a diagonal matrix with $\sum_i|\theta_{ii}|\le 1$, 
in which case $\theta \in G_d$ (see example \ref{ex34}).
 Then ${\cal Q}=|{\rm Tr}(\theta VW^\dagger )|\le 1$ (it cannot take values in the region $(1,k_G)$).
 \end{proposition}
 \begin{proof}
A direct and easy way to prove this is from the following inequalities (for diagonal matrix $\theta$):
\begin{eqnarray}
&&{\cal C}=\left |\sum_{i,j}\theta _{ij}s_it_j\right |\le \sum_{i}|\theta _{ii}|\le 1;\;\;\;|s_i|\le 1;\;\;\;|t_j|\le 1\nonumber\\
&&{\cal Q}=\left |\sum_{i,j}\theta_{ij}\lambda_i\mu_j\bra{u_i}v_j\rangle \right |\le \sum_{i}|\theta _{ii}|\le 1;\;\;\;\lambda_i, \mu_j\le1.
\end{eqnarray}
\end{proof}
It follows that if ${\cal Q}=|{\rm Tr}(\theta VW^\dagger )|\in (1,k_G)$, then the  matrix $\theta$ has non-zero off-diagonal elements, {\bf in a basis such that the assumptions
$W,V\in {\cal S}_d$ and $\theta\in G_d$ hold}.
If the matrix $\theta$  is diagonalisable with a unitary transformation $U$
as $\theta=U\theta_{\rm diag}U^\dagger$ (where $\theta_{\rm diag}$ is a diagonal matrix with its eigenvalues), then
\begin{eqnarray}
{\cal Q}=|{\rm Tr}(\theta VW^\dagger )|=|{\rm Tr}[\theta _{\rm diag}(U^\dagger VU)(U^\dagger W^\dagger U)]|\in(1,k_G).
\end{eqnarray}
There is no contradiction between the right hand side of this equation and our statement  that if ${\cal Q}\in (1,k_G)$ then $\theta$ has non-zero off-diagonal elements .
If $V,W \in {\cal S}_d$,  the 
$U^\dagger VU$ and $U^\dagger W^\dagger U$ might not belong to ${\cal S}_d$ (related to this is Eq.(\ref{9})) and then proposition \ref{GRO} 
does not apply to the right hand side of this expression.

In the `Grothendieck formalism', the assumption ${\cal C}\le 1$ in Eq.(\ref{892})  might not hold after unitary transformations of the matrix $\theta$ (related is Eq.(\ref{341})).
In the formulation in proposition \ref{GRO}, the assumptions $V,W \in {\cal S}_d$ might not hold after unitary transformations.

If $\theta$ is a density matrix, the non-diagonal elements are physically important  because they are related to the superposition principle.
A diagonal density matrix simply represents a probabilistic mixture of states. `Quantumness' is in the non-diagonal elements.
In this sense quantumness (non-diagonal elements in the density matrix) are needed for ${\cal Q}\in(1,k_G)$.

\subsection{Examples and physical importance}

Many physical quantities can be written as ${\rm Tr}(\theta VW^\dagger )$.
The requirement in proposition \ref{GRO} that $\theta \in G_d$ and $W,V\in {\cal S}_d$, is not a restriction, because we have shown that
arbitrary matrices with appropriate normalisation will satisfy this requirement (propositions \ref{pro00}, \ref{pro98}).

Examples:
\begin{itemize}
\item
If $\theta$ is a density matrix and $VW^\dagger$ is an observable (Hermitian operator) then $|{\rm Tr}(\theta VW^\dagger )|$
is the expectation value $|<VW^\dagger>|$.
Alternatively, if $\theta$ is a Hermitian operator and $VW^\dagger$ is a density matrix, then $|{\rm Tr}(\theta VW^\dagger )|$
is the expectation value $|<\theta>|$.
\item
If $\theta$ is a density matrix and $VW^\dagger$ is a displacement operator or displaced parity operator 
(defined for example in \cite{VOU1,VOU} and in the references therein), then 
$|{\rm Tr}(\theta VW^\dagger )|$ is the absolute value of the Weyl or the Wigner function correspondingly.

\end{itemize}

From a physical point of view, the Grothendieck theorem replaces the complex numbers $s_i, t_j\in {\cal D}$ in ${\cal C}$, with the vectors 
$\lambda_i\ket{u_i}$, $\mu_j\ket{v_j}\in{\cal B}_d$ in ${\cal Q}$.  We regard the $s_i, t_j$ as classical quantities, and the $\lambda_i\ket{u_i}$, $\mu_j\ket{v_j}$ 
as the corresponding quantum quantities (which are vectors so that we can have superpositions). Roughly speaking the Grothendieck theorem says that when a classical quadratic form 
${\cal C}$ takes values less than $1$, the corresponding 
quantum quantity ${\cal Q}$ might take values greater than $1$, up to the Grothendieck constant $k_G$.
Quantum Mechanics is described with Hilbert spaces, and should agree with all predictions of the Hilbert space formalism, like the 
Grothendieck bounds.

 In this paper we are particularly interested in the Grothendieck region $(1,k_G)$ which cannot be reached by the corresponding classical models, 
 and which (as we explain below) is related to non-diagonal elements in the density 
 matrix, that in turn are related to the superposition principle.
Quantities for which ${\cal Q}$ takes values in the Grothendieck region $(1,k_G)$, are in the `mathematical ceiling' of the Hilbert space formalism and 
therefore of the quantum formalism that is based on it.
They are quantum mechanics `on the edge'.

We explain below in section \ref{PPP} that important quantities like the Wigner and Weyl functions which show clearly quantum phenomena (like quantum interference),
do not take values in the Grothendieck region.
In this sense the Grothendieck region is a `new territory' with rare examples of quantum mechanics. The term `rare' needs further work to be quantified,
but a result in this direction is in the next section, where we show that there are large families of examples which cannot take values in $(1,k_G)$.

Later in sections \ref{secA2}, \ref{secA3} we give physical examples which take values in $(1,k_G)$.

\section{Families of examples which cannot take values in $(1,k_G)$}\label{PPP}

In the rest of the paper we are interested in the case where ${\cal Q}=|{\rm Tr}(\theta VW^\dagger )|$ with $\theta \in G_d$ and $W,V\in {\cal S}_d$, takes values in the Grothendieck region $(1,k_G)$. 
In this section we show that there are large  families of examples where $\cal Q$ cannot take values in the Grothendieck region $(1, k_G)$.
They include physical quantities like the Wigner and Weyl functions that show clearly quantum interference.

This implies that examples that take values in the Grothendieck region are infrequent.
Especially if we want to get close to $k_G$ we need large matrices \cite{GR5}.
How infrequent examples with values  in the Grothendieck region are, could be quantified in future work by producing randomly a triplet of matrices (normalise them appropriately 
so that the first belongs in $G_d$ and the other two in ${\cal S}_d$) and then find the percentage for which ${\cal Q}\in (1,k_G)$.

The following proposition shows that for normal matrices in $G_d^\prime$ (which can be written as $\frac{K}{de_{\rm max}}$ where $K$ is any normal matrix), we get
${\cal Q}\le 1$. 

\begin{proposition}\label{pro38}
Let $K$ be a $d\times d$ normal matrix, $M$ any $d\times d$ complex matrix, and
\begin{eqnarray}
\theta =\frac{K}{de_{\rm max}}\in G_d ^\prime;\;\;\;V=\frac{M}{{\cal N}(M)}\in{\cal S}_d;\;\;\;W={\bf 1}\in{\cal S}_d.
\end{eqnarray}
Then 
\begin{eqnarray}\label{900A}
{\cal Q}=|{\rm Tr}(\theta VW^\dagger )|=\frac{1}{de_{\rm max}{\cal N}(M)}|{\rm Tr}(K M)|\le 1.
\end{eqnarray}
We get ${\cal Q}=1$ when
$K_{ij}=aM_{ij}$ with $a\le 1$, and also $||K||_2= \sqrt{d}e_{\rm max}$ and $||M||_2= \sqrt {d}{\cal N}(M)$.

\end{proposition}
\begin{proof}
${\rm Tr}(KM)$ can be viewed as inner product of a vector with $d^2$ elements $K_{ij}$ with a vector with $d^2$ elements $M_{ji}$.
We apply the Cauchy-Schwartz inequality and we get:
\begin{eqnarray}
|{\rm Tr}(K M)|=|\sum_{i,j}K_{ij}M_{ji}|\le \sqrt{\sum_{i,j}|K_{ij}|^2}\sqrt{\sum_{i,j}|M_{ji}|^2}=||K||_2||M||_2.
\end{eqnarray}
But for normal matrices, $||K||_2\le \sqrt{d}e_{\rm max}$ (Eq.(\ref{BB1})). Also $||M||_2\le \sqrt {d}{\cal N}(M)$ (Eq.(\ref{8})).
Therefore $||K||_2||M||_2\le de_{\rm max}{\cal N}(M)$. This proves the statement.

In order to get equality in Eq.(\ref{900A}), we need $K_{ij}$ to be proportional to $M_{ij}$ and also 
$||K||_2= \sqrt{d}e_{\rm max}$ and $||M||_2= \sqrt {d}{\cal N}(M)$.
We have seen in Eq.(\ref{BB1})  that for normal matrices ${\cal N}(M)\le e_{\rm max}$ and this gives $a\le 1$.

\end{proof}
\begin{cor}

For a normal matrix $\theta$, a necessary (but not sufficient) condition for ${\cal Q}\in (1,k_G)$ is that $\theta \in G_d\setminus G_d^\prime$.

\end{cor}

\begin{example}
For unitary matrices $e_{\rm max}=1$ and ${\cal N}=1$. We take
\begin{eqnarray}
\theta =\frac{1}{d}F^2\in G_d ^\prime;\;\;\;V=F^2\in{\cal S}_d;\;\;\;W={\bf 1}\in{\cal S}_d
\end{eqnarray}
where $F$ is the Fourier transform in Eq.(\ref{4}), and we get
\begin{eqnarray}
{\cal Q}=|{\rm Tr}(\theta VW^\dagger )|=\frac{1}{d}{\rm Tr}({\bf 1})= 1.
\end{eqnarray}

\end{example}

Below are more cases which cannot take values in the Grothendieck region.

\begin{example}

For a density matrix $\rho$ and a unitary matrix $U$, 
\begin{eqnarray}\label{57}
|{\rm Tr}(\rho U)|\le 1.
\end{eqnarray}
Indeed if $e_i$ and $\ket{e_i}$ are the eigenvalues and eigenvectors of $\rho$, we get
\begin{eqnarray}\label{57}
|{\rm Tr}(\rho U)|=|\sum _ie_i\bra{e_i}U\ket{e_i}|\le \sum_ie_i|\bra{e_i}U\ket{e_i}|\le \sum_ie_i=1.
\end{eqnarray}

Application of the Grothendieck theorem would be to use proposition \ref{GRO} with 
\begin{eqnarray}
\theta=\frac{\rho}{de_{\rm max}}\in G_d^\prime;\;\;\;V=U\in {\cal S}_d;\;\;\;W={\bf 1}\in{\cal S}_d,
\end{eqnarray}
where $e_{\rm max}$ is the spectral radius of $\rho$.
This gives the upper bound 
\begin{eqnarray}\label{RRR}
\frac{1}{de_{\rm max}}|{\rm Tr}(\rho U)|\le \min\left (k_G, \frac{||\rho||_1}{de_{\rm max}}\right )\;\rightarrow\;|{\rm Tr}(\rho U)|\le \min\left (de_{\rm max}k_G, {||\rho||_1}\right )
\end{eqnarray}
We have seen in example \ref{ex39} that $||\rho||_1\ge 1$ and  $de_{\rm max}\ge 1$.
Therefore for $|{\rm Tr}(\rho U)|$, the upper bound $1$ in Eq.(\ref{57}) is better (lower) than the bound in Eq.(\ref{RRR}).
It is seen that $|{\rm Tr}(\rho U)|$ cannot take values in the Grothendieck region which in this case is 
 $de_{\rm max}\times(1,k_G)$.

Physical examples of ${\rm Tr}(\rho U)$ are the Weyl and Wigner functions. 
If $U$ is one of the displacement operators $D(a,b)$
(defined for example in \cite{VOU1,VOU} and in the references therein), then
\begin{eqnarray}
{\cal Q}(a,b)=|{\rm Tr}[\rho D(a,b)]|
\end{eqnarray}
is the absolute value of the Weyl function.
Also if $U$ is one of the displaced parity operators $P(a,b)$ (defined for example in \cite{VOU1,VOU} and in the references therein)  we get the absolute value of the Wigner function
\begin{eqnarray}
{\cal Q}(a,b)=|{\rm Tr}[\rho P(a,b)]|.
\end{eqnarray}
 These functions show quantum behaviour, and yet they cannot take values in the Grothendieck region.
\end{example}

\section{Example in $H(6)$ with values in the Grothendieck  region $(1,k_G)$}\label{secA2}

We give an example that involves physical quantities in a quantum system with $6$-dimensional Hilbert space.
We start with novel generalised coherent states in $H(3)$ with a discrete isotropy. Their overlaps lead to a $6\times 6$ projector $\Pi$. In section \ref{sec3} we 
show that the discrete isotropy of the coherent states leads to the strict inequality $g(\Pi)<6$. Consequently, ${\cal Q}$ that involves ${\lambda}\Pi$ with $\lambda \in (\frac{1}{6},\frac{1}{g(\Pi)})$ 
takes values in the Grothendieck region $(1,k_G)$.

In section \ref{sec4}, we perform a numerical maximisation which proves that $g(\Pi)=5$ and then ${\cal Q}$ can reach the value $\frac{6}{5}$.
If go to large matrices, numerical maximisation with large number of variables can be difficult, and there is merit in analytical methods like the one in section \ref{sec3}.

\subsection{Generalised coherent states in $H(3)$ with discrete isotropy}\label{sec3}
In $H(3)$, we introduce the following $6$ quantum states
\begin{eqnarray}\label{vec}
&&\ket{{a_0}}=\frac{1}{\sqrt{2}}
\begin{pmatrix}
1 \\1\\ 0
\end{pmatrix};\;\;\;
\ket{{a_1}}=\frac{1}{\sqrt{2}}
\begin{pmatrix}
1 \\0\\ 1
\end{pmatrix};\;\;\;
\ket{{a_2}}=\frac{1}{\sqrt{2}}
\begin{pmatrix}
0 \\1\\ 1
\end{pmatrix}\nonumber\\
&&\ket{{a_3}}=\frac{1}{\sqrt{2}}
\begin{pmatrix}
1 \\-1\\ 0
\end{pmatrix};\;\;\;
\ket{{a_4}}=\frac{1}{\sqrt{2}}
\begin{pmatrix}
1 \\0\\ -1
\end{pmatrix};\;\;\;
\ket{a_5}=\frac{1}{\sqrt{2}}
\begin{pmatrix}
0 \\1\\ -1
\end{pmatrix}
\end{eqnarray}
They have been used in ref\cite{GR5} in a real Hilbert space in connection with the real Grothendieck constant.
Here, firstly we interpret them as generalised coherent states in $H(3)$, and we use them in a complex Hilbert space in connection with the complex Grothendieck constant.
We note a discrete isotropy in these states, in the sense that the relationship of a state $\ket{a_i}$ to the rest of them, is the same for all $i$.
This intuitive and qualitative statement is quantified below.

The states $\{\ket{a_i}\}$ are generalised coherent states in the sense of the following proposition:
\begin{proposition}
\begin{itemize}
\item[(1)]
The states $\{\ket{a_i}\}$ resolve the identity:
\begin{eqnarray}\label{r1}
\frac{1}{2}\sum_{i=0}^5\ket{a_i}\bra{a_i}={\bf 1}.
\end{eqnarray}

\item[(2)]
The set $\{\ket{a_0},...,\ket{a_5}\}$ is invariant under transformations in the symmetric group $\Sigma_3$, in the sense that 
for any $\tau_\pi\in \Sigma_3$ (defined in Eq.(\ref{PPQ})) and any $i$ the state $\tau _\pi\ket{a_i}$ is one of the states $\ket{a_j}$ (possibly with a physically insignificant phase factor).
\item[(3)]
There is a `discrete isotropy' between the states $\ket{a_i}$, in the sense that the set of $6$ non-independent probabilities
\begin{eqnarray}
A_i=\{|\bra{a_i}a_j\rangle|^2\;|\;j=0,...5\},
\end{eqnarray}
is the same for all $i$.
Therefore the following sum does not depend on $i$:
 \begin{eqnarray}\label{73}
\sum_{j=0}^5|\bra{a_i}a_j\rangle|^r=1+\frac{1}{2^{r-2}};\;\;\;r=1,2,3,....
\end{eqnarray}
For $r=1$ this sum involves square roots of the probabilities $|\bra{a_i}a_j\rangle|^2$.
For $r=2$ this result follows immediately from the resolution of the identity in Eq.(\ref{r1}).

\end{itemize}
\end{proposition}
\begin{proof}
\begin{itemize}
\item[(1)]
This is proved with direct calculation using the vectors in Eq.(\ref{vec}).
\item[(2)]
This is also proved with direct calculation using the vectors in Eq.(\ref{vec}).
\item[(3)]
We see this through the projector $\Pi$ in Eq.(\ref{64}) below (which contains the overlaps of the coherent states).
Each row or each column has one $|\Pi_{ij}|=\frac{2}{4}$, four $|\Pi_{ij}|=\frac{1}{4}$, and one $|\Pi_{ij}|=0$.
From this follows immediately Eq.(\ref{73}).

\end{itemize}
\end{proof}
Coherent states resolve the identity and their set is invariant under some group of transformations.
Many of the coherent states studied in the literature, are non-orthogonal to each other.
In the present case the state $\ket{a_i}$ is orthogonal to the state $\ket{a_{i+3}}$ (the indices are defined modulo $6$).

For any state $\ket{f}$ the $\{|\bra{f}a_i\rangle|^2\}$ is a set of $6$ non-independent probabilities with 
\begin{eqnarray}
\frac{1}{2}\sum_{i=0}^5|\langle f\ket{a_i}|^2=1.
\end{eqnarray}
The $\ket{a_i}\bra{a_i}$ and $\ket{a_j}\bra{a_j}$ do not commute, and therefore these probabilities are not simultaneously measurable.
They can be measured using different ensembles of the state $\ket{f}$ for each measurement.

An arbitrary state in $H(3)$
\begin{eqnarray}
\ket{f}=\begin{pmatrix}
f_0\\
f_1\\
f_2
\end{pmatrix};\;\;\;|f_0|^2+|f_1|^2+|f_2|^2=1,
\end{eqnarray}
can be expanded in terms of them as
\begin{eqnarray}
\ket{f}=\sum_{i=0}^5 {\mathfrak f}_i \ket{a_i};\;\;\;{\mathfrak f}_i=\frac{1}{2}\bra {a_i} f \rangle;\;\;\;\langle g\ket{f}=\sum _{i=0}^2g_i^*f_i=2\sum _{i=0}^5 {\mathfrak g}_i^*{\mathfrak f}_i.
\end{eqnarray}
It is easily seen that
\begin{eqnarray}\label{ty}
&&{\mathfrak f}_0=\frac{1}{2\sqrt{2}}(f_0+f_1);\;\;\;{\mathfrak f}_1=\frac{1}{2\sqrt{2}}(f_0+f_2);\;\;\;{\mathfrak f}_2=\frac{1}{2\sqrt{2}}(f_1+f_2)\nonumber\\
&&{\mathfrak f}_3=\frac{1}{2\sqrt{2}}(f_0-f_1);\;\;\;{\mathfrak f}_4=\frac{1}{2\sqrt{2}}(f_0-f_2);\;\;\;{\mathfrak f}_5=\frac{1}{2\sqrt{2}}(f_1-f_2).
\end{eqnarray}
and therefore
\begin{eqnarray}\label{nor}
\sum_{i=0}^2|f_i|^2=2\sum_{i=0}^5|{\mathfrak f}_i|^2.
\end{eqnarray}

Since the $\{\ket{a_i}\}$ form an over-complete basis, this expansion is not unique.
It is seen that a quantum state $\ket{f}$ in $H(3)$ can be represented with a $6$-tuple, but 
not every $6$-tuple represents a quantum state in $H(3)$.

\subsection{The projector $\Pi$ of overlaps between the coherent states}

We now consider the $6\times 6$ projector
\begin{eqnarray}\label{64}
\Pi_{ij}=\frac{1}{2}\bra{a_i}a_j\rangle=\frac{1}{4}
\begin{pmatrix}
2&1&1&0&1&1\\
1&2&1&1&0&-1\\
1&1&2&-1&-1&0\\
0&1&-1&2&1&-1\\
1&0&-1&1&2&1\\
1&-1&0&-1&1&2
\end{pmatrix};\;\;\;\Pi^2=\Pi;\;\;\;{\rm rank}(\Pi)=3.
\end{eqnarray}
$\Pi$ has the eigenvalues $1$ (with multiplicity $3$) and $0$ (with multiplicity $3$).

All columns (and rows) of $\Pi$ are $6$-tuples that span a $3$-dimensional space isomorphic to $H(3)$.
But we also have the $3$-dimensional null space of $\Pi$ which we denote as $H(3)_{\rm null}$.
The projector ${\bf 1}-\Pi$ projects into $H(3)_{\rm null}$.
We also define the space 
\begin{eqnarray}
H(6)=H(3)\oplus H(3)_{\rm null}.
\end{eqnarray}

For $\Pi$ we get $de_{\rm max}=6$, and from Eq.({27}) it follows that $g(\Pi)\le 6$. 
We recall here that the proof of this is based on the Cauchy-Schwartz inequality.
We show below that due to the discrete isotropy of the coherent states, the Cauchy-Schwartz inequality cannot become equality in the present context.

\begin{lemma}\label{lem12}
The strict inequality $g(\Pi)< 6$ holds.
\end{lemma}
\begin{proof}
We get equality in Eq.(\ref{A25}) only if Eq.(\ref{A26}) holds (in which case the Cauchy-Schwartz inequality becomes equality). We assume that
the $6$-tuples 
\begin{eqnarray}
&&{\bf s}=(s_0,...,s_5);\;\;\;|s_i|\le 1\nonumber\\
&&{\bf t}=(t_0,...,t_5);\;\;\;|t_i|\le 1,
\end{eqnarray}
give $g(\Pi)=6$. Then ${\bf s}$ and ${\bf t}^\prime=\Pi{\bf t}$ are proportional to each other, and therefore they both belong in $H(3)$.
Since
\begin{eqnarray}
\left |\sum _{i,j}\Pi_{ij}s_it_j\right |=\left |\sum _{i=0}^5s_i t_i^\prime\right |\le \max \left (\sum_{i=0}^5|s_i|^2\right ),
\end{eqnarray}
it follows that a necessary condition for $g(\Pi)=6$ is that all $|s_i|=1$.

Since ${\bf s}$ is in $H(3)$, it can be represented with a vector $(\sigma_0, \sigma_1, \sigma_2)$ (which is not normalised) such that the analogue of Eq.(\ref{ty}) gives 
\begin{eqnarray}\label{564}
&&|\sigma_0+\sigma_1|=2\sqrt{2}|s_0|= 2\sqrt{2};\;\;\;|\sigma_0+\sigma_2|=2\sqrt{2}|s_1|= 2\sqrt{2};\;\;\;|\sigma_1+\sigma_2|=2\sqrt{2}|s_2|= 2\sqrt{2}\nonumber\\
&&|\sigma_0-\sigma_1|=2\sqrt{2}|s_3|= 2\sqrt{2};\;\;\;|\sigma_0-\sigma_2|=2\sqrt{2}|s_4|= 2\sqrt{2};\;\;\;|\sigma_1-\sigma_2|=2\sqrt{2}|s_5|= 2\sqrt{2}
\end{eqnarray}
From this follows easily that
\begin{eqnarray}\label{B}
|\sigma_0|^2+|\sigma_1|^2= 8;\;\;\;|\sigma_0|^2+|\sigma_2|^2= 8;\;\;\;|\sigma_1|^2+|\sigma_2|^2= 8.
\end{eqnarray}
We cannot have equality in all three of these relations,
because this would require 
\begin{eqnarray}\label{B11}
|\sigma_0|=|\sigma_1|=|\sigma_2|=2,
\end{eqnarray}
and this cannot satisfy all relations in Eq.(\ref{564}). 

We note that the sets of equations (\ref{564}), (\ref{B}), (\ref{B11}) are invariant under permutations of the $(\sigma_0, \sigma_1, \sigma_2)$.
This is related to the discrete isotropy of the coherent states that underpin the projector $\Pi$ and the present proof.

We conclude that we cannot have all $|s_i|=1$ and therefore we cannot have $g(\Pi)=6$.

\end{proof}

\subsection{Example in $H(6)$ that involves the projector $\Pi$ and takes values in the Grothendieck  region $(1,k_G)$}\label{sec45}

We consider a quantum system with Hilbert space $H(6)$, and the matrices
\begin{eqnarray}\label{107}
\theta =\lambda \Pi;\;\;\;V=W=\sqrt{2}\Pi\in{\cal S}_6
\end{eqnarray}
Here ${\cal N}(\Pi)=\frac{1}{\sqrt{2}}$ and therefore $\sqrt{2}\Pi\in{\cal S}_6$.
Since $g(\Pi)<6$, we define
\begin{eqnarray}
\epsilon=\frac{1}{g(\Pi)}-\frac{1}{6}>0.
\end{eqnarray}
Therefore Eq(\ref{43}) gives
\begin{eqnarray}
&&{\lambda}\Pi\in G_6\setminus G_6^\prime\;\;\;{\rm for}\;\;\frac{1}{ 6}\le \lambda\le \frac{1}{6}+\epsilon\nonumber\\
&&{\lambda}\Pi\in G_6^\prime\;\;\;{\rm for}\;\;\lambda\le \frac{1}{ 6}.
\end{eqnarray}
For the matrices in Eq.(\ref{107}) we get
\begin{eqnarray}\label{456}
|{\rm Tr}(\theta VW^\dagger )|=2\lambda{\rm Tr}(\Pi)= 6\lambda.
\end{eqnarray}
For $\lambda\le \frac{1}{6}$ (in which case $\theta={\lambda}\Pi\in G_6^\prime$), the  $|{\rm Tr}(\theta VW^\dagger )|$
 takes values in the region$(0,1)$.
 For  $\frac{1}{ 6}\le \lambda\le \frac{1}{6}+\epsilon$ (in which case $\theta ={\lambda}\Pi\in G_6\setminus G_6^\prime$) the  $|{\rm Tr}(\theta VW^\dagger )|$ takes values 
 in  $[1,1+6\epsilon]$ within the Grothendieck region.

We note that $\rho=\frac{1}{6}VW^\dagger =\frac{1}{3}\Pi$ is a density matrix in quantum systems with Hilbert space $H(6)$, and that
\begin{eqnarray}
{\rm Tr}(\rho^2)=\frac{1}{3};\;\;\;E=-{\rm Tr}\rho \log \rho=\log3.
\end{eqnarray}
Also $\Theta=6 \theta=6\lambda \Pi$ is an observable in quantum systems with Hilbert space $H(6)$.
Therefore Eq.(\ref{456}) can be written as
\begin{eqnarray}
|<\Theta>|=|{\rm Tr}(\rho \Theta)|=6\lambda.
\end{eqnarray}
This is an experimentally measurable quantity.

The above analytical method is based on the inequality $g(\Pi)<6$, and it shows that ${\cal Q}$ exceeds $1$.
In the next subsection we perform a numerical maximisation and show that $g(\Pi)=5$.
This gives a maximum value of ${\cal Q}$ equal to $\frac{6}{5}$.

\subsection{Numerical evaluation of $g(\Pi)$}\label{sec4}

In this section we perform numerically the maximisation and find $g(\Pi)=5$. We then use the $\theta =\frac{1}{5}\Pi$ in the Grothendieck theorem.

Since
\begin{eqnarray}
\left |\sum_{i,j}\Pi _{ij}s_it_j\right |\le \frac{1}{2}\max_{\{t_j\}}||\sum _j t_j\ket{a_j}||^2\;\;\;\;|s_i|\le 1;\;\;\;|t_j|\le 1,
\end{eqnarray}
we have to prove that
\begin{eqnarray}
&& \max_{\{t_j\}}||\sum _j t_j\ket{a_j}||^2=\frac{1}{2}(A^2+B^2+C^2)=10;\;\;\;|t_j|\le 1\nonumber\\
 &&A=|t_0+t_1+t_3+t_4|\le 4;\;\;\;B=|t_0+t_2-t_3+t_5|\le 4 ;\;\;\;C=|t_1+t_2-t_4-t_5|\le 4
\end{eqnarray}
We note  that the $t_3, t_4, t_5$ appear with both signs plus and minus in these expressions. Consequently, for
\begin{eqnarray}
&&t_0=t_1=t_2=t_3=t_4=t_5=1,\nonumber\\
&&t_0=t_1=t_2=-t_3=t_4=t_5=1,\nonumber\\
&&t_0=t_1=t_2=t_3=t_4=-t_5=1,
\end{eqnarray}
etc., we get $\{A,B,C\}=\{4,2,0\}$ and $||\sum _j t_j\ket{a_j}||^2=10$.
In order to show that this is the maximum, we take
$t_i=R_i\exp(i\chi_i)$
and we find
\begin{eqnarray}
&&A=|R_0\exp(i\chi_0)+R_1\exp(i\chi_1)+R_3\exp(i\chi_3)+R_4\exp(i\chi_4)|\nonumber\\
&&B=|R_0\exp(i\chi_0)+R_2\exp(i\chi_2)-R_3\exp(i\chi_3)+R_5\exp(i\chi_5)|\nonumber\\
&&C=|R_1\exp(i\chi_1)+R_2\exp(i\chi_2)-R_4\exp(i\chi_4)-R_5\exp(i\chi_5)|.
\end{eqnarray}
We used the MATLAB algorithm `fmincon' to minimise $-(A^2+B^2+C^2)$ which is a function of the $12$ variables $(R_i,\chi_i)$ 
with the constraints 
\begin{eqnarray}
R_i\in[0,1];\;\;\;\chi_i\in [-\pi,\pi].
\end{eqnarray}
This is a gradient based algorithm, which starts from an initial value of the $12$ parameters and moves towards the minimum.
We used many initial values of the parameters in the neighbourhood of $t_0=t_1=t_2=t_3=t_4=t_5=1$ (e.g., $R_i=0.5, \chi_i=0.5$) and they all showed that the minimum 
is at $R_i=1$ and $\chi_i=0$.
This shows that $ \max_{\{t_j\}}||\sum _j t_j\ket{a_j}||^2=10$ and therefore $g(\Pi)=5$.

Using this with the example in section \ref{sec45}, we get a maximum value of ${\cal Q}$ equal to $\frac{6}{5}$.

\section{Example in $H(12)$ with values in the Grothendieck  region $(1,k_G)$}\label{secA3}

This example is a generalisation of the previous one.
We use the same notation, so that we can avoid repetition as much as possible.

We first note that the vectors in Eq.(\ref{vec}) can be viewed as the columns of the Fourier matrix (with $d=2$),
\begin{eqnarray}
F=\frac{1}{\sqrt{2}}
\begin{pmatrix}
1&1\\
1&-1
\end{pmatrix}
\end{eqnarray}
 `diluted' with zeros. Inspired by this observation, we do below something similar with the Fourier matrix for $d=3$:
 \begin{eqnarray}
F=\frac{1}{\sqrt{3}}
\begin{pmatrix}
1&1&1\\
1&\omega&\omega^2\\
1&\omega^2&\omega
\end{pmatrix};\;\;\;\omega=\exp\left (i\frac{2\pi}{3}\right ).
\end{eqnarray}
 
 \subsection{Generalised coherent states in $H(4)$ with discrete isotropy}\label{sec3}  
In a $H(4)$ we consider the $12$ normalised vectors $\ket{a_i}$ in table \ref{t1} (the components of $\sqrt{3}\ket{a_i}$ are shown). Here $\omega=\exp(i\frac{2\pi}{3})$.
The states $\{\ket{a_i}\}$ are generalised coherent states in the sense of the following proposition:
\begin{proposition}
\begin{itemize}
\item[(1)]
The states $\{\ket{a_i}\}$ resolve the identity:
\begin{eqnarray}\label{r2}
\frac{1}{3}\sum_{i=0}^{11}\ket{a_i}\bra{a_i}={\bf 1}.
\end{eqnarray}
\item[(2)]
The set $\{\ket{a_0},...,\ket{a_{11}}\}$ is invariant under transformations in the symmetric group $\Sigma_4$, in the sense that 
for any $\tau_\pi\in \Sigma_4$ (defined in Eq.(\ref{PPQ})) and any $i$ the state $\tau _\pi\ket{a_i}$ is one of the states $\ket{a_j}$ (possibly with a physically insignificant phase factor).
\item[(3)]
There is a `discrete isotropy' between the states $\ket{a_i}$, in the sense that the set of $12$ non-independent probabilities
\begin{eqnarray}
A_i=\{|\bra{a_i}a_j\rangle|^2\;|\;j=0,...11\},
\end{eqnarray}
is the same for all $i$.
Therefore the following sum does not depend on $i$:
 \begin{eqnarray}\label{730}
\sum_{j=0}^{11}|\bra{a_i}a_j\rangle|^r=1+\frac{2^r+2}{3^{r-1}};\;\;\;r=1,2,3,....
\end{eqnarray}
For $r=1$ this sum involves square roots of the probabilities $|\bra{a_i}a_j\rangle|^2$.
For $r=2$ this result follows immediately from the resolution of the identity in Eq.(\ref{r2}).
\end{itemize}
\end{proposition}
\begin{proof}
\begin{itemize}
\item[(1)]
This is proved with direct calculation using the vectors in table \ref{t1}.
\item[(2)]
This is also proved with direct calculation using the vectors in table \ref{t1}.
\item[(3)]
This is seen in the projector in table \ref{t2} below (which is related to $\bra{a_i}a_j\rangle$).
In each row or in each column of  we have one $|\Pi_{ij}|=\frac{3}{9}$, three $|\Pi_{ij}|=\frac{2}{9}$, six $|\Pi_{ij}|=\frac{1}{9}$, and two $|\Pi_{ij}|=0$.
From this follows Eq.(\ref{730}).
\end{itemize}
\end{proof}

An arbitrary state in $H(4)$
\begin{eqnarray}
\ket{f}=\begin{pmatrix}
f_0\\
f_1\\
f_2\\
f_3
\end{pmatrix}
\end{eqnarray}
can be expanded in terms of them as
\begin{eqnarray}
\ket{f}=\sum_{i=0}^{11} {\mathfrak f}_i \ket{a_i};\;\;\;{\mathfrak f}_i=\frac{1}{3}\bra {a_i} f \rangle;\;\;\;\langle g\ket{f}=\sum_{i=0}^3g_i^*f_i=3\sum _{i=0}^{11} {\mathfrak g}_i^*{\mathfrak f}_i.
\end{eqnarray}
From this we get $12$ relations
\begin{eqnarray}\label{ty12}
{\mathfrak f}_0=\frac{1}{3\sqrt{3}}(f_0+f_1+f_2);\;\;\;{\mathfrak f}_1=\frac{1}{3\sqrt{3}}(f_0+\omega^2f_1+\omega f_2);\;\;\;{\mathfrak f}_2=\frac{1}{3\sqrt{3}}(f_0+\omega f_1+\omega^2f_2);\;\;\;{\rm e.t.c.}
\end{eqnarray}
and we show that
\begin{eqnarray}\label{nor12}
\sum_{i=0}^3|f_i|^2=3\sum_{i=0}^{11}|{\mathfrak f}_i|^2.
\end{eqnarray}

\subsection{The projector $\Pi$ of overlaps between the coherent states}
We  introduce the $12\times 12$ projector $\Pi_{ij}=\frac{1}{3}\bra{a_i}a_j\rangle$ given in table \ref{t2}
(the matrix elements of $9\Pi_{ij}$ are shown).
This matrix has the eigenvalues $1$ (with multiplicity $4$) and $0$ (with multiplicity $8$).

All columns (and rows) of $\Pi$ are $12$-tuples that span a $4$-dimensional space isomorphic to $H(4)$.
But we also have the $8$-dimensional null space of $\Pi$ which we denote as $H(8)_{\rm null}$.
The projector ${\bf 1}-\Pi$ projects into $H(8)_{\rm null}$.
We define the space 
\begin{eqnarray}
H(12)=H(4)\oplus H(8)_{\rm null}.
\end{eqnarray}
For $\Pi$ we get $de_{\rm max}=12$, and from Eq.({27}) it follows that $g(\Pi)\le 12$.

\begin{lemma}
The strict inequality $g(\Pi)< 12$ holds.
\end{lemma}
\begin{proof}
The proof is similar to the proof of lemma \ref{lem12}, and we present briefly the differences.
The gist of the proof is that due to the discrete isotropy of the coherent states, the Cauchy-Schwartz inequality cannot become equality in the present context.

We assume that the $12$-tuples ${\bf s}, {\bf t}$ (with $|s_i|\le 1$ and $|t_i|\le 1$) give $g(\Pi)=12$.
Then ${\bf s}$ and ${\bf t}^\prime=\Pi{\bf t}$ are proportional to each other, and therefore they both belong in $H(4)$.
Since
\begin{eqnarray}
\left |\sum _{i,j}\Pi_{ij}s_it_j\right |=\left |\sum _{i=0}^{11}s_i t_i^\prime\right |\le \max \left (\sum_{i=0}^{11}|s_i|^2\right ),
\end{eqnarray}
it follows that a necessary condition for $g(\Pi)=12$ is that all $|s_i|=1$.

Since ${\bf s}$ is in $H(4)$, it can be represented with a vector $(\sigma_0, \sigma_1, \sigma_2, \sigma_3)$ (which is not normalised) such that the analogue of Eq.(\ref{ty12}) gives 
$12$ inequalities:
\begin{eqnarray}\label{564A}
&&|\sigma_0+\sigma_1+\sigma_2|= 3\sqrt{3}|s_0|=3\sqrt{3};\;\;\;|\sigma_0+\omega^2 \sigma_1+\omega \sigma_2|= 3\sqrt{3}|s_1|=3\sqrt{3}\nonumber\\
&&|\sigma_0+\omega \sigma_1+\omega^2 \sigma_2|= 3\sqrt{3}|s_2|=3\sqrt{3};\;\;\;{\rm e.t.c.}
\end{eqnarray}
We note that the set of these equations is invariant under permutations of the $(\sigma_0, \sigma_1, \sigma_2, \sigma_3)$.
We add the squares all these $12$ inequalities.
Using the fact that $1+\omega+\omega^2=0$, we find that all terms $\sigma_i \sigma_j^*$ with $i\ne j$, cancel. Therefore we get
\begin{eqnarray}\label{PP11}
\sum_{i=0}^3|\sigma_i|^2= 36.
\end{eqnarray}
Due to symmetry, equality in this relation would require that all $|\sigma_i|=3$.
Then from Eqs.(\ref{564A}) we get the $12$ equations
\begin{eqnarray}
&&\sigma_0\sigma_1^*+\sigma_1\sigma_2^*+\sigma_2\sigma_0^*+\sigma_0^*\sigma_1+\sigma_1^*\sigma_2+\sigma_2^*\sigma_0=0\nonumber\\
&&\sigma_0\sigma_1^*\omega+\sigma_0^*\sigma_1\omega^2+\sigma_2\sigma_0^*\omega+\sigma_0\sigma_2^*\omega^2+\sigma_1\sigma_2^*\omega+\sigma_2\sigma_1^*\omega^2=0\nonumber\\
&&\sigma_0\sigma_1^*\omega^2+\sigma_0^*\sigma_1\omega +\sigma_2\sigma_0^*\omega^2+\sigma_0\sigma_2^*\omega+\sigma_1^*\sigma_2\omega+\sigma_2^*\sigma_1\omega^2=0;\;\;\;{\rm e.t.c.}
\end{eqnarray}
We multiply the first three equations by $1, \omega^2, \omega$ correspondingly, and adding them we get:
\begin{eqnarray}\label{111}
\sigma_0\sigma_1^*+\sigma_1\sigma_2^*+\sigma_2\sigma_0^*=0.
\end{eqnarray}
In a similar way we also get
\begin{eqnarray}
\sigma_0\sigma_1^*+\sigma_1\sigma_3^*+\sigma_3\sigma_0^*=0;\;\;\;
\sigma_0\sigma_3^*+\sigma_3\sigma_2^*+\sigma_2\sigma_0^*=0;\;\;\;
\sigma_1\sigma_2^*+\sigma_2\sigma_3^*+\sigma_3\sigma_1^*=0.
\end{eqnarray}
We can find solutions for each of these equations. For example
if $\sigma_i=3\exp(i\phi_i)$, Eq.(\ref{111}) has the solution $\phi_0-\phi_1=\frac{2\pi}{3}$ and $\phi_0=\phi_2$ (and its permutations).
But we cannot satisfy simultaneously all these four equations.

We conclude that we cannot have all $|s_i|=1$ and therefore we cannot have $g(\Pi)=12$.
Throughout the proof we have a symmetry under permutations of the $(\sigma_0, \sigma_1, \sigma_2, \sigma_3)$.
This is related to the discrete isotropy of the coherent states.

\end{proof}

\subsection{Example in $H(12)$ that involves the projector $\Pi$ and takes values in the Grothendieck  region $(1,k_G)$}

Since $g(\Pi)<12$, we define
\begin{eqnarray}
\epsilon=\frac{1}{g(\Pi)}-\frac{1}{12}>0.
\end{eqnarray}
and then
\begin{eqnarray}
&&{\lambda}\Pi\in G_{12}\setminus G_{12}^\prime\;\;\;{\rm for}\;\;\frac{1}{ 12}\le \lambda\le \frac{1}{12}+\epsilon\nonumber\\
&&{\lambda}\Pi\in G_{12}^\prime\;\;\;{\rm for}\;\;\lambda\le \frac{1}{ 12}.
\end{eqnarray}

We then apply this to a physical example analogous to the one in section \ref{sec45}  (here $\sqrt{3}\Pi\in{\cal S}_{12}$) and we show that for $\frac{1}{ 12}\le \lambda\le \frac{1}{12}+\epsilon$ we get ${\cal Q}$ within the Grothendieck region $(1,k_G)$.

We do not perform a numerical maximisation in this example (it involves $24$ variables).

\begin{remark}
Both examples in sections 7,8 involve overcomplete sets of states with a resolution of the identity (in Eqs(\ref{r1}), (\ref{r2})).
Overcomplete sets of states with resolution of the identity have been used extensively in the context of coherent states, but we note that here some of these states are orthogonal to each other (there are zeros in the two projectors).
These two examples are interesting in their own right, because they might bring a novel aspect in the general area of phase space methods for systems with finite-dimensional Hilbert space.

We also note that the projectors (Eq.(\ref{64}) and table \ref{t2}) can be linked to reproducing kernels ($\sum \Pi_{ij}{\mathfrak f}_j={\mathfrak f}_i$).
The reproducing kernel formalism has been used extensively in maximisation problems (e.g., in the context of Machine Learning and Artificial Inteligence\cite{A1,A2,A3}), and this might provide a 
deeper understanding of why these two examples lead to values within the Grothendieck region, which is on the edge of the Hilbert space formalism and consequently of the Quantum formalism.
\end{remark}

\section{Discussion}
We have used the  Grothendieck theorem in the context of a single quantum system, in contrast to previous work that used it
in the context of multipartite entangled systems. 
In this paper:
\begin{itemize}
\item
We have reformulated the  Grothendieck theorem in terms of products of normalised matrices so that it can be used with measurable quantum mechanical quantities.
 The normalisation factors and their properties have been discussed in sections 3,4.
\item
We are particularly interested in the Grothendieck region $(1,k_G)$, which is a kind of `quantum ceiling'.
We have shown that there are large  families of examples  with $\cal Q$ that cannot take values in the Grothendieck region $(1, k_G)$.
We also gave necessary (but not sufficient) conditions  for ${\cal Q}$ taking values in the Grothendieck region
(corollary \ref{GRO10} and proposition \ref{pro38}). 
In Mathematics sometimes the interesting features are found in rare examples on the `edge' (which here is the Grothendieck bound $k_G$).
Section 6 indicates that examples with ${\cal Q}\in(1,k_G)$ might be rare.
Further work with random sampling of matrices, can quantify how rare are the examples with ${\cal Q}\in (1, k_G)$.

\item 
A physical example in the Hilbert space $H(6)$, that leads to  ${\cal Q}=\frac{6}{5}$ has been discussed in section 7.
We presented both an analytical method which shows that ${\cal Q}$ enters the Grothendieck region $(1,k_G)$, 
and a numerical method that gives the maximum value ${\cal Q}=\frac{6}{5}$.
A similar  example in the Hilbert space $H(12)$ has been discussed in section 8 (the analytical approach only).

These two examples use $d(d-1)$ novel generalised coherent states in $H(d)$, which resolve the identity and have a discrete isotropy (with $d=3,4$).
Their overlaps define a $(d^2-d)\times (d^2-d)$ projector. We considered its null space $H(d^2-2d)_{\rm null}$ and the space
$H(d^2-d)=H(d)\oplus H(d^2-2d)_{\rm null}$. We then proved that  for $d=3,4$ we get the strict inequality $g(\Pi)<d^2-d$ and this led to values in the Grothendieck region.
These examples might be generalised into $d\ge 5$.

More work is needed on examples that take values very close to  $k(d)\le k_G\le 1.4049$. The present work indicates that they might be related to generalised coherent states with discrete isotropy as symmetry.
Helpful in this direction is the link between coherent states with the reproducing formalism and in turn with maximisation problems.
Such examples illuminate a new territory at the edge of quantum mechanics.
Another related problem is to find better estimates of $k(d)$ and for $k_G$ and improve the results in \cite{GR5,GR6,GR7}.
\end{itemize}
The work explores the Grothendieck theorem in the context of a single quantum system, with emphasis on 
the Grothendieck region $(1,k_G)$ which is on the edge of the Hilbert space formalism and the quantum formalism.

\newpage
\begin{table}
\caption{$12$ vectors $\ket{a_i}$ in $H(4)$. The components of $\sqrt{3}\ket{a_i}$ are shown. Here $\omega=\exp(i\frac{2\pi}{3})$. }
\def\arraystretch{2}
\begin{tabular}{|c|c|c|c|c|c|c|c|c|c|c|c|}\hline
$\sqrt{3}\ket{a_0}$&$\sqrt{3}\ket{a_1}$&$\sqrt{3}\ket{a_2}$&$\sqrt{3}\ket{a_3}$&$\sqrt{3}\ket{a_4}$&$\sqrt{3}\ket{a_5}$&$\sqrt{3}\ket{a_6}$&$\sqrt{3}\ket{a_7}$&$\sqrt{3}\ket{a_8}$&$\sqrt{3}\ket{a_9}$&$\sqrt{3}\ket{a_{10}}$&$\sqrt{3}\ket{a_{11}}$\\\hline
$1$&$1$&$1$&$0$&$0$&$0$&$1$&$\omega^2$&$\omega$&$1$&$\omega$&$\omega^2$\\\hline
$1$&$\omega$&$\omega^2$&$1$&$1$&$1$&$0$&$0$&$0$&$1$&$\omega^2$&$\omega$\\\hline
$1$&$\omega^2$&$\omega$&$1$&$\omega$&$\omega^2$&$1$&$1$&$1$&$0$&$0$&$0$\\\hline
$0$&$0$&$0$&$1$&$\omega^2$&$\omega$&$1$&$\omega$&$\omega^2$&$1$&$1$&$1$\\\hline
\end{tabular} \label{t1}
\end{table}

\begin{table}
\caption{ The projector $\Pi_{ij}=\frac{1}{3}\bra{a_i}a_j\rangle $ where ${\ket{a_i}}$ are the vectors in table \ref{t1}.
The matrix elements of $9\Pi_{ij}$  are shown. Here $\omega=\exp(i\frac{2\pi}{3})$. }
\def\arraystretch{2}
\begin{tabular}{|c|c|c|c|c|c|c|c|c|c|c|c|}\hline
$3$&$0$&$0$&$2$&$-\omega^2$&$-\omega$&$2$&$-\omega$&$-\omega^2$&$2$&$-1$&$-1$\\\hline
$0$&$3$&$0$&$-1$&$2\omega^2$&$-\omega$&$-\omega^2$&$-1$&$2\omega$&$-\omega$&$2\omega$&$-\omega$\\\hline
$0$&$0$&$3$&$-1$&$-\omega^2$&$2\omega$&$-\omega$&$2\omega^2$&$-1$&$-\omega^2$&$-\omega^2$&$2\omega^2$\\\hline
$2$&$-1$&$-1$&$3$&$0$&$0$&$2$&$-\omega^2$&$-\omega$&$2$&$-\omega$&$-\omega^2$\\\hline
$-\omega$&$2\omega$&$-\omega$&$0$&$3$&$0$&$-1$&$2\omega^2$&$-\omega$&$-\omega^2$&$-1$&$2\omega$\\\hline
$-\omega^2$&$-\omega^2$&$2\omega^2$&$0$&$0$&$3$&$-1$&$-\omega^2$&$2\omega$&$-\omega$&$2\omega^2$&$-1$\\\hline
$2$&$-\omega$&$-\omega^2$&$2$&$-1$&$-1$&$3$&$0$&$0$&$2$&$-\omega^2$&$-\omega$\\\hline
$-\omega^2$&$-1$&$2\omega$&$-\omega$&$2\omega$&$-\omega$&$0$&$3$&$0$&$-1$&$2\omega^2$&$-\omega$\\\hline
$-\omega$&$2\omega^2$&$-1$&$-\omega^2$&$-\omega^2$&$2\omega^2$&$0$&$0$&$3$&$-1$&$-\omega^2$&$2\omega$\\\hline
$2$&$-\omega^2$&$-\omega$&$2$&$-\omega$&$-\omega^2$&$2$&$-1$&$-1$&$3$&$0$&$0$\\\hline
$-1$&$2\omega^2$&$-\omega$&$-\omega^2$&$-1$&$2\omega$&$-\omega$&$2\omega$&$-\omega$&$0$&$3$&$0$\\\hline
$-1$&$-\omega^2$&$2\omega$&$-\omega$&$2\omega^2$&$-1$&$-\omega^2$&$-\omega^2$&$2\omega^2$&$0$&$0$&$3$\\\hline
\end{tabular} \label{t2}
\end{table}

\end{document}